%Paper: astro-ph/9311055
%From: TPSINGH@tifrvax.tifr.res.in
%Date: Mon, 22 Nov 1993 19:04 +0530

%Plain Tex

\magnification\magstep 1
\headline={\ifnum\pageno=1\hfil\else\hfil\tenrm--\ \folio\ --\hfil\fi}
\footline={\hfil}
\hsize=6.0truein
\vsize=8.54truein
\hoffset=0.25truein
\voffset=0.25truein
\baselineskip=12pt
\parskip=0.2 cm
\parindent=1cm
\tolerance=10000
\pretolerance=10000

\def\d{\hbox {d}}
\def\etal{{et al.~}}
\def\mg{\big <}
\def\md{\big >}
\def\dta{\delta}
\def\i{\hbox{i}}
\def \vx{\bf x}
\def\init{\tabskip 0pt }
\def\crr{\cr \noalign{\hrule}}

\rightline {November, 1993}
\vglue 2 true in
\centerline{\bf{THE QUASI-LINEAR EVOLUTION OF THE DENSITY FIELD}}
\centerline{\bf{IN MODELS OF GRAVITATIONAL INSTABILITY}}\medskip
\centerline{$\rm F.~Bernardeau^1,~ T. ~ P.~Singh^2,~ B.~Banerjee^2,
            {}~ S. ~M. ~Chitre^2$}
\smallskip
\centerline {$\rm ^1 Canadian~ Institute~ for~ Theoretical~ Astrophysics$}
\centerline{University of Toronto, Toronto, Ontario, Canada M5S 1A1}
\centerline{$\rm ^2 Tata~ Institute~ of~ Fundamental~ Research$}
\centerline{Homi Bhabha Road, Bombay 400 005, India}\medskip
\vskip 1 in
\noindent{\bf ABSTRACT}
\medskip
\noindent Two quasi-linear approximations, the
frozen flow approximation (FFA) and the frozen potential approximation (FPA),
have been proposed recently
for studying the evolution of a collisionless
self-gravitating fluid. In the FFA it is assumed that the velocity
field remains unchanged from  its value obtained from the linear
theory whereas in FPA the same approximation is made
for the gravitational potential. In this paper we compare these
and the older Zel'dovich approximation by calculating the
evolution of the density in perturbation theory. In particular we compute
the skewness, including the smoothing effects,
and the kurtosis for the FFA, FPA and Zel'dovich
 approximation and compare their relative accuracy.
\medskip
\noindent {\bf Keywords:} galaxies: clustering $-$
cosmology: theory $-$ large-scale
structure of Universe.
\vfill\eject

\noindent {\bf 1. INTRODUCTION}\smallskip
\noindent Large--scale structures observed today are believed to
have developed from small density fluctuations generated in the
early universe. The growth of these fluctuations has been studied
by regarding the system to be a collisionless self-gravitating
fluid. When the amplitudes of the fluctuations are small, perturbation
theory can be use to study the evolution of the system. In particular
the growth rate of the rms fluctuations can be described by the
linearised equations of  fluid motion.
When the fluctuations eventually become nonlinear, however,
the perturbation theory is no more accurate
and N-body simulations have been widely used to overcome this
difficulty. The understanding of the physical processes that take
place in such a self-gravitating fluid necessitates, however,
the use of analytical models and approximations that can be more
easily studied.

The best known model for describing the mildly nonlinear evolution
is due to Zel'dovich (1970) (also see Zel'dovich \& Shandarin
1989). In this approximation the motion of each particle is determined
by its initial Lagrangian displacement. A good presentation of
this approximation has been made by Moutarde \etal 1989 and by
Bouchet \etal 1992 in the frame of a Lagrangian description.
The evolution of the density field
can be studied in this approximation until the formation of
caustics, when this approach breaks down. Recently two new approximations
 have been proposed, with the aim of improving upon
the Zel'dovich approximation -- (1) the frozen flow approximation (FFA)
 (Matarrese \etal 1992) and (2) the frozen potential approximation (FPA)
 (Brainerd, Scherrer \& Villumsen 1993, Bagla \& Padmanabhan 1993).
 In FFA the
velocity flows are `frozen' to their local initial linear values, and
at any time the velocity of each particle is the one associated to the
point at which it lies. The evolution of the density is then treated exactly.
In the second approximation, FPA, the gravitational potential
is `frozen' at its linear value. That is, the Eulerian potential
is kept constant and the particles obey the standard Eulerian
equations of motion in this potential.

All the three approximations mentioned above are naturally
consistent with the linear theory. However beyond the linear
order the density
evolution is different in each case. To compare them we
calculate, assuming Gaussian initial conditions,
the third and fourth order moments of density by
means of perturbation theory. We then compare the results obtained
for each of these approximations with those
 obtained from perturbation theory using the exact dynamics.

In Sec. 2 we recall the basic  equations of motion and the
calculations of third and fourth  moments of density (Peebles 1980,
Fry 1984, Grinstein \& Wise 1987, Bouchet \etal 1992, Bernardeau 1993)
for the exact dynamics as well as for the
Zel'dovich approximation. These computations are repeated for FFA and FPA
in Sec. 3 and 4 respectively.
Throughout we assume an $\Omega =1$ spatially flat universe,
so that the scale factor evolves as
$a(t)=a_0(t/t_0)^{2/3}$.

\vskip 3 cm
\noindent{\bf 2. BASIC EQUATIONS}\smallskip
\noindent The evolution of a collisionless self-gravitating
fluid in an expanding Robertson-Walker universe
is described by the following equations in the Newtonian
limit (Peebles 1980),

$$\dot\delta +{1\over a}
{\bf\nabla}.\left[(1+\delta){\bf v}\right]=0,\eqno(1)$$
$$\dot{\bf v}+
{\dot a\over a}{\bf v}+
{1\over a}
\left[{\bf v}.{\bf\nabla}\right]
{\bf v}=-{1\over a}
{\bf\nabla}\phi,\eqno(2)$$
$${\bf\nabla}^2\phi=4\pi G\rho_b a^2\delta.\eqno(3)$$

Here, \noindent $\delta\equiv[\rho({\bf x},t)$
$-\rho_b(t)]/\rho_b(t)$ is the
enhancement of the true density
$\rho({\bf x},t)$ over the mean density
$\rho_b(t)$, {\bf v} is the proper
peculiar velocity, relative
to the Hubble flow and $\phi({\bf x},t)$ is the peculiar
gravitational potential.

{}From these equations it follows that the density contrast
$\delta({\bf x},t)$ evolves according to the
equation
$$\eqalign{\ddot\delta+
{2\dot a\over a}\dot\delta-
4\pi G\rho_b\delta&=
4\pi G\rho_b\delta^2+
{1\over a^2}
{\bf\nabla}_i\delta.{\bf\nabla}_i\phi\cr
&+{1\over a^2}{\bf\nabla}_i{\bf\nabla}_j
\left[(1+\delta)v^iv^j\right].\cr}\eqno(4)$$

In the linear theory, all the terms on the right hand side
of (4) are dropped, and $\delta\equiv \delta^{(1)}$
$({\bf x},t)$ has the solution $\delta^{(1)}({\bf x},t)$
$=A({\bf x})D(t)$, with
$D(t)\propto a(t)\propto t^{2/3}$
(growing mode) for $\Omega=1$.

It is convenient to define the potential $\Delta({\bf x},t)$
through the relation $\phi=4\pi G\rho_ba^2\Delta$, so that
$\nabla^2\Delta=\delta$. The peculiar velocity
${\bf v}^{(1)}$
in the linear theory is
$${\bf v}^{(1)}=-{a\dot D\over D}
{\bf {\bf\nabla}}\Delta^{(1)},\eqno(5)$$
where
$$\Delta^{(1)}({\bf x})=-
{1\over 4\pi}\int d^3 x^{\prime}
{\delta({\bf x}^{\prime})\over
|{\bf x-x}^{\prime}|}
={\phi^{(1)}\over 4\pi G\rho_b a^2}.\eqno(6)$$
The solution for $\delta$ in perturbation theory is obtained
via the perturbation expansion
$$\delta=\sum^{\infty}_{n=1}
\delta^{(n)},
{\bf v}=\sum^{\infty}_{n=1}
{\bf v}^{(n)},
\Delta=\sum^{\infty}_{n=1}\Delta^{(n)},\eqno(7)$$
with respect to the initial Gaussian
fluctuations so that $\delta^{(n)}$ satisfies the equation
(Fry 1984)
$$\eqalign{ \ddot\delta^{(n)}&+
{2\dot a\over a}\dot\delta^{(n)}-4\pi G\rho_b\delta^{(n)}=
\sum^{n-1}_{k=1}
\left[4\pi G\rho_b\delta^{(k)}
\delta^{(n-k)}\right.\cr
&+\left.4\pi G\rho_b{\bf\nabla}_i\delta^{(k)}
{\bf\nabla}_i\Delta^{(n-k)}+
{1\over a^2}{\bf\nabla}_i{\bf\nabla}_j
v^{(k)i}v^{(n-k)j}\right.\cr
&+\left. {1\over a^2}\sum^{k-1}_{m=1}
{\bf\nabla}_i{\bf\nabla}_j
\delta^{(m)}v^{(k-m)i}v^{(n-k)j}\right].\cr}\eqno(8)$$

The derivation of the behavior of the first cumulants
of the density distribution at large scale can be done assuming
Gaussian initial conditions. It can be shown that in general
(Goroff \etal 1986, Bernardeau 1992)
$$\mg\dta^p\md_c=S_p\mg\dta^2\md^{p-1}.\eqno(9)$$
where $S_p$ is a coefficient that depends weakly on the cosmological
parameters. Moreover when the smoothing effects are neglected
these coefficients are independent of the shape of the
power spectrum. Bernardeau (1992) gives a method to derive
the whole series of the coefficients for the exact dynamics, but we
consider here only the first two coefficients $S_3$ and $S_4$.

Let us first recall the principle of the calculation of $S_3$ and
$S_4$ for the exact dynamics. The derivation of the skewness of the
distribution function requires the calculation of the
density contrast at second
order, $\delta^{(2)}$. It is obtained
by setting $n=2$ in equation (8),
$$\eqalign{ \ddot\delta^{(2)}+
{2\dot a\over a}
\dot\delta^{(2)}-
4\pi G\rho_b\delta^{(2)}&=
4\pi G\rho_b
\left(\delta^{(1)2}+
{\bf\nabla}_i\delta^{(1)}
{\bf\nabla}_i\Delta^{(1)}\right)\cr
&+{1\over a^2}
{\bf\nabla}_i{\bf\nabla}_j
\left[v^{(1)i}v^{(1)j}\right].\cr}$$
The solution for the growing mode is
$$\delta^{(2)}={5\over 7}
\delta^{(1)2}+
\delta^{(1)}_{,i}
\Delta^{(1)}_{,i}+
{2\over 7}\Delta^{(1)2}_{,ij}.\eqno(10)$$
Using this solution, it is straightforward to
calculate the skewness $S_3\equiv\mg\delta^3\md/\mg\delta^2\md^2$, which
to the lowest order is
$3\mg\delta^{(1)2}\delta^{(2)}\md/\mg\delta^{(1)2}\md^2=34/7$
(Peebles 1980).

The calculation of the fourth cumulant involves the knowledge of the
density field at the third order in perturbative calculation. It can
be shown that the coefficient $S_4$ can be written
(Fry 1984, Bernardeau 1992) as $S_4=12R_a+4R_b$, where
$$R_a=4\mg\delta^{(1)2}\delta^{(2)2}\md
/\mg\dta^{(1)2}\md^3,\eqno(11)$$
and
$$R_b=\mg\delta^{(1)3}\delta^{(3)}\md
/\mg\dta^{(1)2}\md^3.\eqno(12)$$
The value of $R_a$ can be easily obtained from Eqn. (10),
$R_a=(34/21)^2=2.62$.
The third order term of the density contrast,
$\delta^{(3)}$, can be calculated more easily in Fourier space,
by first defining
$$\tilde{\delta}({\bf k})={1\over V}
\int \d^3\vx\delta({\bf x})e^{\i{\bf k.x}},\eqno(13)$$
and similar transforms for ${\bf v}$ and $\Delta$. Eqn. (8) for
$n=3$ then gives a solution for $\tilde{ \delta}^{(3)}$,
which can be used to show that $R_b=682/189=3.61$, and that
$S_4=45.88$ (see Fry 1984, Bernardeau 1992).

These results, however, concern the behaviour of the cumulants
at a given point. When the field is filtered at a given scale, the values of
$S_3$ and $S_4$ have to be changed. We recall here the results obtained for
a top hat window function (Juszkiewicz, Bouchet \& Colombi 1993,
Bernardeau 1993).
They read,
$$S_3={34\over 7}+\gamma_1\eqno(14)$$
and
$$S_4={60712\over 1323}+{62\over 3}\gamma_1+
{7\over 3}\gamma_1^2+{2\over 3}\gamma_2\eqno(15)$$
where $\gamma_1$ and $\gamma_2$ are the first two logarithmic derivatives
of the variance with scale,
$$\eqalignno{
\gamma_1&={\d\log\mg\delta^2\md\over \d\log R}&(16),\cr
\gamma_2&={\d^2\log\mg\delta^2\md\over \d\log^2 R}.&(17)\cr}$$
The derivation of these smoothing effects is based on geometrical properties
of the top hat window function given by Bernardeau (1993).

For the Zel'dovich approximation the behaviour of the cumulants at large
scale is similar to the one encountered in the real dynamics but the
coefficients $S_p$ are slightly changed due to the approximation
that is made (Grinstein \& Wise 1987). Recently Bernardeau (1993)
has derived the expression of the coefficients $S_3$ and $S_4$
when a top hat filter is applied to the density field,
$$\eqalignno{
S_3^{Zel}&=4+\gamma_1,&(18)\cr
S_4^{Zel}&={272\over9}+{50\over3}\gamma_1+
{7\over 3}\gamma_1^2+{2\over 3}\gamma_2.&(19)\cr}$$

This quantitative change of the large--scale cumulants is general for
any approximative dynamics starting with Gaussian initial conditions.
The coefficients $S_p$ then
turn out to be a good tool to test the various approximations
with each other by comparing the values of these coefficients.
In the next parts we will generalize the results obtained for
the Zel'dovich approximation to the frozen flow and the
frozen potential approximations.

\vskip 3 cm
\noindent{\bf 3. FROZEN FLOW APPROXIMATION}\smallskip
\noindent The frozen flow approximation (FFA) which was
proposed by Matarrese \etal (1992) is best
defined using slightly different variables from those in
Eqns. (1-3). Using the scale factor `a' as the time variable, define
the comoving peculiar velocity ${\bf u}\equiv d{\bf x}/da$
$={\bf v}/a\dot a$.
$\eta=1+\delta$,
$\psi\equiv (3t^2_0/2a^3_0)\phi$, where
$a(t)=a_0(t/t_0)^{2/3}$. Eqns. (1-3) then reduce to
$${d\eta\over da}+\eta {\bf\nabla}.{\bf u}=0,\eqno(20)$$
$${d{\bf u}\over da}+
{3\over 2a}{\bf u}=-
{3\over 2a}{\bf\nabla}\psi,\eqno(21)$$
$$\nabla^2\psi=\delta/a,\eqno(22)$$
where $d/da=\partial/\partial a+{\bf u.}{\bf\nabla}$.

FFA is defined by assuming that the velocity field
${\bf u}$ is steady:
$\partial{\bf u}/\partial a=0$; that is, stream lines
are frozen to their initial shape. The frozen
value of ${\bf u}$ would then be the constant value
it has in the linear theory:
$${\bf u}_{FFA}({\bf x})=-
{\bf\nabla}\psi_{LIN}({\bf x}).\eqno(23)$$
(This of course implies that ${\bf v=v}_{LIN}={\bf v}^{(1)}$,
as given by Eqn. (5).) The ${\bf u}_{FFA}$
of Eqn. (23) is a solution of the Euler equation
(21) provided $\psi$ is approximated to be
$$\psi_{FFA}({\bf x},t)=
\psi_{LIN}-
{a\over 3}
({\bf\nabla}\psi_{LIN})^2.\eqno(24)$$
In the notation of Eqns. (1-3), FFA corresponds to
$$\eqalign{ {\bf v}_{FFA}=
{\bf v}^{(1)},
\phi_{FFA}&=
\phi^{(1)}-{a\over 3}
\left({3 t^2_0\over 2a^3_0}\right)
\left({\bf\nabla}\phi^{(1)}\right)^2\cr
&=\phi^{(1)}-{4\pi\over 3}
G\rho_b a^2\left({\bf\nabla}\Delta^{(1)}\right)^2\cr
&\equiv\phi^{(1)}+\phi^{(2)}.   \cr } \eqno (25)$$
The form of $\phi^{(2)}$ implies that it is second order
in perturbation, but it is not the same as the second
order potential in true non-linear evolution. To study
the density evolution in FFA, we first note that
Eqns. (1-3) give
$$\eqalign { \ddot \delta +{2 \dot a \over a}\dot\delta &=
{1\over a^2}
(1+\delta)
\nabla^2\phi+{1\over a^2}
{\bf\nabla}\delta. {\bf\nabla}\phi\cr
&+ {1\over a^2}{\bf\nabla}_i
{\bf\nabla}_j
\left[(1+\delta)v^iv^j \right ]. \cr } \eqno (26)$$
In here, we substitute for ${\bf v}$ as ${\bf v}_{FFA}$
and for $\phi$ as $\phi_{FFA}$. Next, we implement the
perturbation expansion of Eqn. (7)
to get the following equation for $\delta^{(2)}$:
$$\eqalign {\ddot \delta^{(2)}+
{2 \dot a \over a} \dot \delta^{(2)}-
{1 \over a^2}
\nabla^2 \phi^{(2)} &=
4 \pi G \rho_b
\delta^{(1)2}+{1 \over a^2}
{\bf\nabla} \delta^{(1)}. {\bf\nabla} \phi^{(1)}\cr
& +{1 \over a^2}{\bf\nabla}_i {\bf\nabla}_j
\left [ v^{(1)i}v^{(1)j} \right ].\cr} \eqno (27)$$
This equation has the solution
$$\delta^{(2)}_{FFA}=
{1 \over 2} \delta^{(1)2}+
{1 \over 2} \delta^{(1)}_{,i} \Delta^{(1)}_{,i} \eqno (28)$$
which should be compared with the $\delta^{(2)}$  for the
true evolution, in Eqn. (10). From here, it is straightforward
to carry through the calculation of skewness as in Peebles (1980),
Section 18, since the only change in the solution $\delta^{(2)}$ is
that in the coefficients. The result is $S_3^{FFA}=3$, as
compared to the true value of 34/7.
The smoothing effects on $S_3$ can be easily calculated and the final
result for a top hat window function reads,
$$S_3^{FFA}=3+{\gamma_1\over 2}.\eqno(29)$$

To obtain $S_4$ in $FFA$, we first find $R_a$ from Eqn. (11) by
substituting the solution $\delta^{(2)}_{FFA}$. This gives, upon
angle averaging, as in Fry (1984), $R_a=1.0$.
{}From (26), the Eqn. for $\delta^{(3)}$ in FFA reads
$$\eqalign{ \ddot\delta^{(3)}+
{2\dot a\over a}
\dot\delta^{(3)}&=
{1\over a^2}
\left\{\delta^{(2)}\nabla^2\phi^{(1)}+\delta^{(1)}
\nabla^2\phi^{(2)}\right\}\cr
&+{1\over a^2}
\left\{{\bf\nabla}\delta^{(2)}.{\bf\nabla}\phi^{(1)}+
{\bf\nabla}\delta^{(1)}.{\bf\nabla}\phi^{(2)}\right\}\cr
&+{1\over a^2}{\bf\nabla}_i{\bf\nabla}_j
\left[\delta^{(1)}v^{(1)i}v^{(1)j}\right].\cr}\eqno(30)$$
Before doing the Fourier transform, we note that
$${\bf\nabla}\phi^{(2)}=-
{8\pi\over 3}G\rho_b a^2
\left({\bf\nabla}\Delta^{(1)}.{\bf\nabla}\right){\bf\nabla}\Delta^{(1)},
\eqno(31)$$
$$\nabla^2\phi^{(2)}=-
{8\pi\over 3} G\rho_ba^2
\left\{\Delta^{(1)}_{,i}\delta^{(1)}_{,i}+
\Delta^{(1)}_{,ij}\Delta^{(1)}_{,ij}\right\}.\eqno(32)$$
Using the transform (13) and its inverse
$$\delta({\bf x})=
{V\over (2\pi)^3}\int
d^3k\tilde{\delta}({\bf k})e^{i{\bf k.x}},\eqno(33)$$
the Fourier transform of a product can be written as a
convolution (Fry, 1984):

$$\eqalign {FT \left \{ F_1(x)\cdots F_N(x)\right\} &=
{1\over V^N}\int
{d^3k_1 \over (2\pi)^3} \cdots
{d^3 k_N \over (2 \pi)^3}
\left [ V \delta_D (\sum {\bf k}_i-{\bf k})\right ] \cr
& \tilde{ F}_1({\bf k}_1)\cdots
\tilde{ F}_N({\bf k}_N)=
\tilde{ F}_1 * \cdots * \tilde{ F}_N.\cr} \eqno (34)$$
When applied to Eqn. (30) for
$\delta^{(3)}_{FFA}$, this gives

$$\eqalign {\ddot { \tilde{ \delta}} ^{(3)}  &+
{2 \dot a \over a} \dot {\tilde{ \delta}} ^{(3)}  =
4 \pi G \rho_b
\left \{ \tilde{ \delta}^{(2)} {*}
\tilde{ \delta}^{(1)}-
{2 \over 3} \tilde{ \delta}^{(1)}*
\left ( {k_i \tilde{ \delta}^{(1)} \over
k^2} \right)*
\left (k_i \tilde{ \delta}^{(1)} \right) \right \} \cr
\qquad \qquad &- {2 \over 3} \tilde{ \delta}^{(1)} *
\left( {k_ik_j \over k^2} \tilde{ \delta}^{(1)} \right) *
\left ( {k_i k_j \over k^2} \tilde{ \delta}^{(1)} \right)\cr
&+ \left( k_i \tilde{ \delta}^{(2)} \right)*
\left( {k_i \over k^2}\tilde{\delta}^{(1)}\right)-{2\over 3}
\left(k_i\tilde{\delta}^{(1)}\right)*
\left({k_j\tilde{\delta}^{(1)}\over
k^2}\right)*
\left({k_ik_j\tilde{\delta}^{(1)}\over
k^2}\right)\cr
&+{1\over a^2}(ik_i)(ik_j)
\tilde{\delta}^{(1)}*
\tilde{ v}^{(1)i}*\tilde{ v}^{(2)j}.\cr}\eqno(35)$$
The solution can be written down in analogy with
Eqn. (46) of Fry (1984), so we do not put
it down explicitly, except to note that now
$\delta^{(2)}_{FFA}$
of Eqn. (28) should be used. Using this $\tilde{\delta}^{(3)}$ and carrying
out angular averaging as in Fry's paper gives $R_b=1$, and hence that
$$S_4^{FFA}=16.\eqno(36)$$
The derivation of the smoothing effects for $S_4$
is slightly more complicated and will not be given.

\vfill\eject
\noindent{\bf 4. FROZEN POTENTIAL APPROXIMATION}\smallskip
\noindent The frozen potential approximation (FPA) was proposed by
Brainerd \etal (1993) and by Bagla \& Padmanabhan (1993).
FPA is defined by keeping the potential $\phi$ constant
at its value $\phi^{(1)}({\bf x})$ in the linear
theory, so that $\phi^{(2)}$ and higher terms in the perturbation expansion
of $\phi$ are set to zero. However, unlike in FFA, ${\bf v}$ is not
approximated,
but is to be obtained from the Euler equation (2) with
$\phi({\bf x},t)\equiv \phi^{(1)}$
$({\bf x})$. Thus Eqn. (26) for the evolution of density, when
written for the case of FPA, becomes
$$\eqalign{\ddot\delta+{2\dot a\over a}\dot\delta&=
{1\over a^2}(1+\delta)
\nabla^2\phi^{(1)}+
{1\over a^2}
{\bf\nabla}\delta.{\bf\nabla}\phi^{(1)}\cr
&+{1\over a^2}{\bf\nabla}_i{\bf\nabla}_j
\left[(1+\delta)v^iv^j\right].\cr}\eqno(37)$$
Using the perturbation expansion (7), the equation
for $\delta^{(2)}$ is found to be
$$\eqalign{ \ddot\delta^{(2)}+{2\dot a\over a}
\dot\delta^{(2)}&=
4\pi G\rho_b\delta^{(1)2}+
{1\over a^2}
{\bf\nabla}\delta^{(1)}.{\bf\nabla}\phi^{(1)}\cr
&+{1\over a^2}{\bf\nabla}_i{\bf\nabla}_j
\left[v^{(1)i}v^{(1)j}\right].\cr}\eqno(38)$$
The solution is
$$\delta^{(2)}={1\over 2}
\delta^{(1)2}+{7\over 10}
\delta^{(1)}_{,i}\Delta^{(1)}_{,i}+
{1\over 5}
\Delta^{(1)2}_{,ij},\eqno(39)$$
as contrasted to the true $\delta^{(2)}$ in Eqn. (10), and
$\delta^{(2)}_{FFA}$
in Eqn. (28). Once again, the skewness is easy to
work out, following Peebles (1980), and the
result is $S^{FPA}_3=17/5$. The derivation of the smoothing effects
gives
$$S_3^{FPA}={17\over 5}+{7\over 10}\gamma_1.\eqno(40)$$

The calculation of the kurtosis of the density field can be
done as usual: Substituting this expression for $\delta^{(2)}$ in (11) and
applying angle averaging as before, gives
$R_a=(17/15)^2=1.28$.
The equation for $\delta^{(3)}$ in FPA is
$$\eqalign{ \ddot\delta^{(3)}+
{2\dot a\over a}\dot\delta^{(3)}&=
4\pi G\rho_b\delta^{(1)}\delta^{(2)}+
{1\over a^2}{\bf\nabla}\delta^{(2)}.{\bf\nabla}\phi^{(1)}\cr
&+{1\over a^2}{\bf\nabla}_i{\bf\nabla}_j
\left[\delta^{(1)}v^{(1)i}v^{(1)j}+
v^{(1)i}v^{(2)j}+v^{(2)i}v^{(1)j}\right],\cr}\eqno(41)$$
as contrasted to the equation (30) for $\delta^{(3)}_{FFA}$.
Here, $\delta^{(2)}$ is the FPA solution Eqn. (39), and the solution for
${\bf v}^{(2)}$ can be found from the
continuity equation to be
$${\bf v}^{(2)}_{FPA}=-
{4a\over 3t}{\bf \nabla}
\Delta^{(2)}_{FPA}-
\delta^{(1)}{\bf v}^{(1)}.\eqno(42)$$
Eqn. (41) is a special case of the equation for the true
$\delta^{(3)}$, and has been obtained simply by setting
$\phi^{(2)}=0$.
It is thus easier to handle than the equation
for $\delta^{(3)}_{FFA}$.
Carrying out the Fourier transform and angular
averaging precisely as in Fry (1984) gives
$R_b=457/315=1.45$, and hence
$$S_4^{FPA}=21.22.\eqno(43)$$

\vskip 1.5 cm
\noindent{\bf 5. CONCLUSIONS}\smallskip
\noindent In Table 1, we display the leading order mean values of
the third and fourth moments for the true and approximate
evolution. For the third moment, the smoothing corrections for a top hat
window function are included, ($n+3=-\d\log\mg\dta^2\md/\d\log R$).
\midinsert
{\baselineskip 14 pt
\centerline{\bf Table 1}
$$\vbox{\init\halign to 14 truecm
{
\strut#&#\tabskip=1em plus 2em&
\hfil$#$\hfil&
$\,$#&
\hfil$#$\hfil&
$\,$#&
\hfil$#$\hfil&
#\tabskip 0pt\crr
&& && {\rm\bf Third\ moment,\ } S_3 && {\rm\bf Fourth\ moment,\ } S_4 &\crr
&&&&&\cr
&& {\rm True\ evolution} && 4.86-(n+3) && 45.88 &\cr
&& {\rm Zel'dovich} && 4.00-(n+3) && 30.22 &\cr
&& {\rm Frozen\ potential} && 3.40-0.7\ (n+3) && 21.22 &\cr
&& {\rm Frozen\ flow} && 3.00-0.5\ (n+3) && 16.00 &\crr
}}$$}
\endinsert

  How do our results compare with those from simulations using N-body,
FFA and FPA? (For results on these simulations, see
Matarrese \etal 1992, Brainerd \etal 1993, Bagla \& Padmanabhan 1993,
Mellott \etal 1993).
At first sight, one might think that since both FFA and
FPA simulations are more similar to N-body than Zel'dovich
approximation, our analytical
results conflict with results of density evolution from the simulations.
However, while FFA and FPA both
prevent the thickening of pancakes, they do not do well in moving mass
to the right place. Brainerd \etal (1993) find from their
analysis that the cross-correlation between N-body and Zel'dovich is
higher than that between N-body and FPA. Also, Mellott \etal (1993) find that
FFA does poorly in cross-correlation with N-body.

At this stage it is useful to compare the second order solution for
the density field in the various approximations. For the Zel'dovich case,
the second order solution is (Bouchet \etal 1992)
$$\delta^{(2)}={1\over 2}
\delta^{(1)2}+
\delta^{(1)}_{,i}\Delta^{(1)}_{,i}+
{1\over 2}
\Delta^{(1)2}_{,ij},\eqno(44)$$
and the second order solution for FFA and FPA is given in Eqns. (28) and (39)
respectively. The same terms appear in the three approximations, but
the coefficients are different.
The coefficients in the Zel'dovich case are the closest to the true density
evolution, followed by FPA and then FFA, and this feature is clearly reflected
in the respective values of the skewness.
It is now interesting to try to understand why these approximations
underestimate such a quantity. The skewness can be seen as a measure
of the ability of the system to create rare dense spots. This interpretation
seems to be confirmed by the relation of the spherical collapse with the
value of the skewness (see Bernardeau 1993). As can be seen in
Matarrese \etal (1992),
the FFA is less accurate than the Zel'dovich
approximation (and the real dynamics)
for the spherical collapse: the acceleration is too
weak to concentrate the matter efficiently.
This result is again borne out by the result on the kurtosis.

When the smoothing effects are taken into
account, however, the results for the skewness seem to attenuate
these effects. In particular for $n=-1$
the Zel'dovich approximation, the FFA and the FPA all give the same result,
$S_3=2$ (instead of the exact value $S_3=2.86$).
Actually the smoothing corrections are sensitive to the tidal effects
in the density field (they come from the term
$\delta^{(1)}_{,i}\Delta^{(1)}_{,i}$) and only the Zel'dovich approximation
gives the right coefficient for this term. The other two approximations
underestimate the tidal effects. The FFA even
fails to give a term containing a quadrupole contribution
(in $\Delta^{(1)}_{,ij}\Delta^{(1)}_{,ij}$). This is a major consideration
for a practical use of these approximations. The disruption
of objects, for instance, is expected to be less accurate in the
FFA or the FPA than in the real dynamics.

   FFA was proposed to improve upon the Zel'dovich approximation, by
avoiding shell-crossing. In a similar spirit, FPA attempts to improve
upon FFA and Zel'dovich, by keeping only the potential linear, but evolving
both density and velocity exactly
(shell-crossing does take place). However, our results suggest that
FFA and FPA are only partially successful in their aim, and the
quasilinear regime is rather poorly approximated by these
approximations.
Thus while the simulations and our analytical results are two different
ways of testing the various approximations, both the means
indicate the need for a more careful comparison, before FFA and FPA
can be adopted as improvements over the Zel'dovich approximation.
These approximations might of course be
interesting for analytical studies, but one should generally exercise caution
in their use.

\medskip
\noindent F.B. and T.P.S. would
like to thank the organizers of the School on Cosmology and Large
Scale Structure, at Les Houches, where some of these ideas were
discussed.
\vfill\eject
\item{} {\bf REFERENCES}
\medskip
\item{} Bagla, J. S. \& Padmanabhan, T., 1993, MNRAS, in press
\item{} Bernardeau, F. 1992, ApJ, 392, 1
\item{} Bernardeau, F., 1993, CITA/93/44 preprint
\item{} Bouchet, F., Juszkiewicz, R., Colombi, S. \& Pellat, R., 1992,
ApJ, 394, L5
\item{} Brainerd, T.G., Scherrer, R.J. \&
Villumsen, J.V., 1993 ApJ, in press
\item{} Fry, J.N., 1984, ApJ, 279, 499
\item{} Grinstein, B. \& Wise, M.B., 1987, ApJ, 320, 448
\item{} Goroff, M.H., Grinstein, B., Rey, S.-J. \& Wise, M.B.,
1986, ApJ, 311, 6
\item{} Juszkiewicz, R., Bouchet, F. \& Colombi, S., 1993, ApJ, 412, L9
\item{} Matarrese, S., Lucchin, F., Moscardini, L. \&
Saez, D., 1992, MNRAS, 259, 437
\item{} Melott, A.L., Lucchin, F., Matarrese, S. \&
Moscardini, L., 1993, preprint, astro-ph/9308008
\item{} Moutarde, F, Alimi J.-M., Bouchet F.R., Pellat, R. \& Ramani,
A., 1991, ApJ, 382, 377
\item{} Peebles, P.J.E., 1980, The Large Scale Structure of the
Universe. Princeton University Press, Princeton, NJ
\item{} Shandarin, S.F. \& Zel'dovich Ya.B., 1989, Rev Mod Phys,
61, 185
\item{} Zel'dovich, Ya. B., 1970, A \& A, 5, 84
\end